\documentclass[%
 reprint,
 amsmath,amssymb,
 aps,
prl,
]{revtex4-1}

\usepackage[utf8]{inputenc}  
\usepackage[T1]{fontenc}
\usepackage{comment}
\usepackage{graphicx}
\usepackage{dcolumn}
\usepackage{bm}

\usepackage[dvipsnames]{xcolor}
\usepackage[colorlinks=true]{hyperref}

\newcommand{\ft}[1]{{\color{black} #1}}
\newcommand{\nw}[1]{{\color{black} #1}}
\newcommand{\Pe}[0]{\mathrm{Pe}}
\newcommand{\rmin}[0]{{r_{\rm min}} }
\begin{document}

\preprint{APS/123-QED}

\title{{Phase separation} and multibody {effects} in three-dimensional active Brownian particles}

\author{Francesco Turci}
\email[Corresponding author: ]{f.turci@bristol.ac.uk}
\author{Nigel B.\ Wilding}
\affiliation{H.H. Wills Physics Laboratory, Tyndall Avenue, Bristol, BS8 1TL, UK}

\begin{abstract}
Simulation studies of the phase diagram of repulsive active Brownian particles in three dimensions reveal that the region of motility-induced phase separation between a high and low density phase is enclosed by a region of gas-crystal phase separation. Near-critical loci and structural crossovers can additionally be identified in analogy with simple fluids. Motivated by the striking similarity to the behaviour of equilibrium fluids with short-ranged pair-wise attractions, we show that a direct mapping to pair potentials in the dilute limit implies interactions that are insufficiently attractive to engender phase separation. Instead, this is driven by the emergence of multi-body effects associated with particle caging that occurs at sufficiently high number density. We quantify these effects via information-theoretical measures of $n$-body effective interactions extracted from the configurational structure.
\end{abstract}

\maketitle

{Liquid-vapor} phase separation and critical behaviour are well-known characteristics of equilibrium fluids which stem from the presence of attractive interactions amongst their atoms or molecules \cite{widom1967,sengers1986}.  Non-equilibrium many-particle systems can also display similar features: for example, assemblies of self-propelled (active) particles undergo -- even in the absence of cohesive forces -- a so-called motility-induced phase separation (MIPS) \cite{cates2015motility,marchetti2016minimal,solon2018} between a high and low-density phase. Manifestations of this phenomenon in the form of aggregation have been observed both in suspensions of motile bacteria \cite{liu2019self} and self-propelled colloids \cite{buttinoni2013dynamical}. 

A key model that captures the essence of the physics of these systems is active Brownian particles (ABPs) \cite{fily2012athermal}, where the constituents interact via pair-wise repulsive forces in the absence of hydrodynamic interactions \cite{marchetti2013}. 
\ft{Despite their nonequilibrium nature, some aspects of ABPs' behaviour can be rationalized in terms of equilibrium concepts such as a pressure equation of state }\nw{by incorporating out-of-equilibrium contributions and modifying equilibrium physical principles, e.g. by using an altered Maxwell construction \cite{speck2014a,winkler2015,fodor2016,levis2017}.}

A frequently highlighted feature of active systems is their ``cooperativity'', characterised in terms of the kinetics of the constituents and the pattern of their orientation in space \cite{bechinger2016,pietzonka2019autonomous}. Even for the simple case of ABPs, which lack an alignment mechanism, cooperative motion has been observed in two (2d) and three dimensions (3d) to give rise to complex patterns in the bulk and to engender phase separation \cite{stenhammar2014a,wysocki2014}. Cooperative motion has also been suggested as a possible cause of the non-monotonic response to {the intensity of self propulsion} -- i.e. the activity -- in the relaxation of active glasses \cite{klongvessa2019}. 

\nw{Particle orientations are important to account for cooperativity for the microscopic explanation of collective effects in active matter \cite{nemoto2019}. However, to describe phase behavior, attempts have been made to construct minimal coarse-grained models that retain only particle coordinates or a continuum density profile, similar to simple models of equilibrium fluids \cite{speck2014a,farage2015,marconi2016,trefz2016,slowman2016,wittmann2017,solon2018}. Within this approach, the influence of \ft{orientational correlations} are subsumed into `effective' inter-particle interaction parameters. }

In this Letter we develop further the effective interaction approach to gain \ft{novel} insight into phase separation and cooperativity. 
We first establish in detail the properties of a 3d system of ABPs including the phase diagram and critical point, together with features of the single phase region that also occur in equilibrium liquids namely a ``Widom line'' of maximum correlation length \cite{xu2005relation,hansen2013};  a line of maximum \ft{number density fluctuations}; and a line of structural crossover - from exponential to oscillatory - known as the Fisher-Widom line \cite{vega1995location,statt2016}. We observe that the phase behaviour is strongly reminiscent of that occurring in equilibrium fluids having a very short-ranged pair potential, with a critical point enclosed by the region of crystal-vapor phase separation and an order parameter broadly consistent with Ising universality. However, a coarse-grained model described by an effective pair potential derived in the low density limit is completely unable to account for MIPS. A quantitative analysis demonstrates, instead, that \nw{non-perturbative} multi-body interactions arise spontaneously in the active system, promoting effective attractions that ultimately drive the phase separation. 

\begin{figure}[t]
    \centering
    \includegraphics{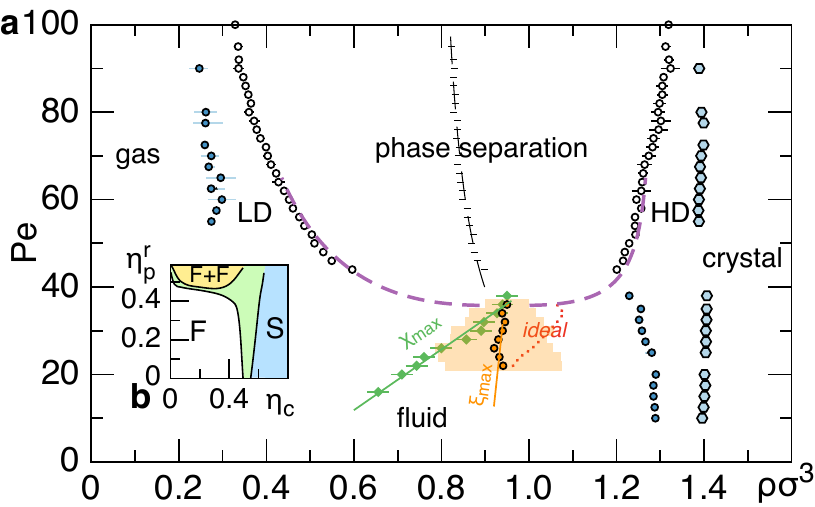}
    \caption{
    (a) Phase diagram of 3d ABPs: empty circles correspond to the binodal of MIPS, separating a low density (LD) from a (HD) fluid; blue circles and hexagons are the fluid (or vapor) and {crystal} branches of fluid-solid coexistence; horizontal dashes are the coexistence diameter, fitted by a rectilinear diameter law $\frac{1}{2}(\rho_{LD}+\rho_{HD})=a \tau+\rho_c$. {Lines of maximal $\xi$ (orange), maximal $\chi/\chi_{ideal}$ (green),  crossover $\chi/\chi_{ideal}=1$ (red dots) are also plotted.} (b) Phase diagram of a colloid-polymer mixture with size ratio $q=0.4$ displaying fluid (F) solid (S) and metastable fluid-fluid coexistence (F+F) regions \ft{parametrized by the colloid and polymer reservoir packing fractions $\eta_c,\eta_p$}, adapted from \cite{dijkstra1999}.}
    \label{fig:phasediagram}
\end{figure}

\ft{Previous simulation work has concentrated on 2d ABPs which display specific features such as a orientationally ordered hexatic phase \cite{digregorio2018,klamser2018}. The present study broadens the discussion to 3d ABPs where different phase behavior is expected \cite{stenhammar2014a,  wysocki2014}. In the model, particles interact via short-ranged, pairwise, repulsive interactions of the Weeks-Chandler-Anderson potential (WCA) form with length-scale $\sigma$, have constant self-propulsion velocity $v_0$ and coupled translational and rotational diffusivities $D_{\rm t}=D_r\sigma^2/3$, so that only the density $\rho$ and the P\'eclet number $\Pe=v_0/(\sigma D_r)$ are the relevant control parameters.} We first determine the phase diagram in the $\rho$-$\Pe$ plane. To do so, the system is initiated in an elongated, periodic simulation box(slab) with a pair of high/low density interfaces. We characterise the  phases that emerge in the steady state. Previous work has focused on the MIPS between two disordered (fluid-like) phases, although studies in 3d also report the presence of an ordered face centred cubic {crystal} at high density and low $\Pe$ \cite{wysocki2014}. We confirm the stability of the crystalline phase at high $\Pe$, and also trace lines of both crystal-vapor coexistence and a MIPS binodal to derive a complete phase diagram, Fig.~\ref{fig:phasediagram}(a).

\begin{figure}[b]
\centering
  \includegraphics[width=\columnwidth]{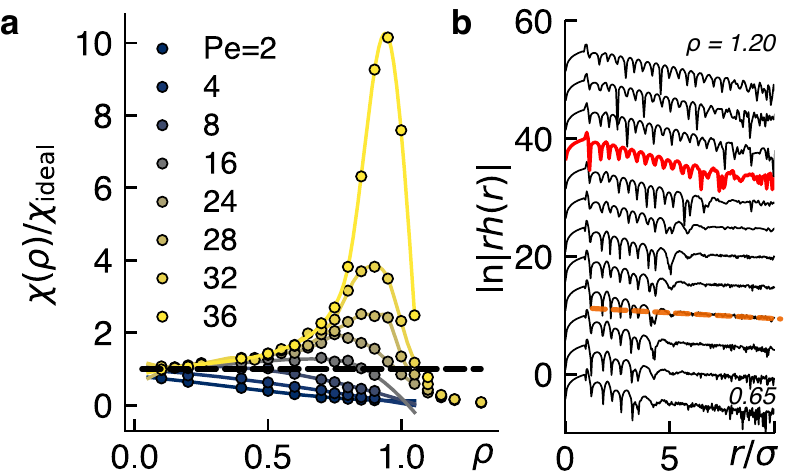}
  \caption{
  (a) Estimates of $\chi/\chi_{\rm ideal}$ versus $\rho$. The horizontal dashed line marks the ideal gas value. (b) Total correlation function $h(r)$ for {Pe=28 measured for a selection of densities in [0.60, 1.20]. A dashed line (orange) shows a representative Ornstein-Zernicke fit to identify the bulk correlation length $\xi$. The red profile corresponds to density $\rho=1.05$, where $\chi/\chi_{ideal}\approx1$ and the exponential decay ceases.} }
  \label{fig:critical}
\end{figure}

The onset of MIPS is associated with critical behaviour \cite{stenhammar2014a} and it is important to characterise the critical point and the near-critical region. In simple liquids exhibiting liquid-vapor (LV) criticality, the single phase region {displays} several crossover lines; in particular, lines of maxima of {compressibility} and correlation length are found which emanate from the critical point and serve as a means to estimate its location. We ask whether analogous features occur for a motility induced critical point. 

A block analysis \cite{rovere1988,villamaina2014} of the particle number fluctuations allows us to measure \nw{a quantity analogous to the relative compressibility of an equilibrium fluid}  $\chi/ \chi_{\rm ideal}:=(\langle N^2\rangle-\langle N\rangle^2)/\langle N\rangle$, while the decay at large distances of the total correlation function $h(r)= g(r)-1$ allows us to determine the bulk correlation length $\xi$, see the Supplemental Material \cite{SI}. Fig.~\ref{fig:critical}(a) shows clearly that for $\Pe \gtrsim 10$  fluctuations grow with increasing $\Pe$. The loci of the maxima of $\chi/ \chi_{\rm ideal}$ and $\xi$ identify two lines in the near-critical single phase region. Within numerical uncertainty, the crossing of the two lines provides a good estimate for the location of the critical point: $\rho_c \approx 0.94$ and $\Pe_c\approx 36$. In several equilibrium fluids \cite{stopper2019decay} the  supercritical density for which $\chi/\chi_{ideal}=1$ closely approximates the so-called ``Fisher-Widom'' line of structural crossover, between a regime with oscillatory decay of the correlations to a regime with purely exponential decay, {see Fig.~\ref{fig:critical}(b)}. This line - Fig.~\ref{fig:phasediagram}(a) - \ft{is located at high densities and points to a structural crossover only close to the fluid-solid transition as} discussed further in \cite{SI}.

Since the P\'eclet number plays a \ft{role akin to the inverse temperature} of equilibrium system, we follow a previous proposal \cite{siebert2018} and define $\tau= |\Pe^{-1}-\Pe_c^{-1}|/\Pe_c^{-1}$ as the reduced P\'eclet number. We can then contrast several properties of MIPS with the behaviour familiar from simple liquids. We find: (i) that the critical point is located at particular high densities, resulting in a rather asymmetric coexistence region (similarly to the 2d case \cite{siebert2018,digregorio2018} ); (ii) that the near-critical region of the binodal can be fitted with Ising forms $\Delta \rho_{i} = A_i \tau^{\beta}$ with $\beta=\beta_{\rm 3dIsing}=0.326$, consistent with recent on-lattice modelling of active particles \cite{partridge2019}; (iii) while the coexistence diameter $d_{\rho}=(\rho_{\rm LD}+\rho_{\rm HD})/2$ does not vary linearly with respect to $\Pe$, it does follow a linear relationship with respect to $\tau$, $d_{\rho}=\rho_c + a \tau$, as in simple equilibrium liquids \cite{cornfeld1972,panagiotopoulos1994}. 

Our phase diagram shows that the MIPS region is enclosed within the region of crystal-vapor coexistence. Indeed, the overall topology of our phase diagram is reminiscent of that of equilibrium fluids having very short ranged pair interactions \cite{miller2004, dijkstra1999}. Here the archetypal system is a colloid ($c$)- polymer ($p$) mixture with size ratio $q=\sigma_p/\sigma_c$. $q$ determines the range of effective colloid-colloid interactions leading to colloid-rich and colloid-poor phase separation. When $q\ll 1$,  the vapour-liquid binodal becomes metastable with respect to crystal-vapor coexistence.
While in principle an effective one-component model requires a multi-body description of the colloid-colloid interactions, in practice a short ranged pair potential accurately accounts for the phase behaviour for sufficiently small $q$ \cite{binder2014,kobayashi2019}.

\nw{Inspired by these similarities, we enquire whether it is possible to construct a coarse-grained effective equilibrium (or {\em passive}) model capable of reproducing the true phase behaviour. An exact
passive model will have inter-particle interactions
yielding the same probability of observing a given particle configuration as
the active model. In general, one expects that this requires an effective Hamiltonian which is sum of $n$-body contributions $H^{\rm eff}=\sum_{n=2}^\infty \theta_n$. Truncating this series at finite $n$ yields an approximate passive model, \ft{e.g. $n=2$ is the $2$-body approximation.}

To gain systematic insight into the role of $n$-body effective interactions in the passive model,  we study systems containing successively larger particle
numbers $N=2, 3, 4, \ldots$ in a box of volume $V$. The $N=2$ system
yields an exact effective pair potential $W_2(r)$ (see below).  Studies
of $N=3$ ABPs provide information on 3-body contributions to the
effective interactions.  Our method quantifies these not in terms of a
3-body interaction potential $\theta_3$ but in terms of the excess free energy of the
{\em exact} effective passive model that describes $N=3$ ABPs, compared to
that of the approximate passive model of $N=3$ particles interacting solely via the pair potential $W_2(r)$.  In a similar way, a study of $N=4$ particles yields information on the $4$-body contribution to the excess free energy unexplained by $2$- and $3$-body interactions. 

Operationally, one computes the 1d probability distribution
function $P_N(\rmin)$ to find a minimal interparticle separation
distance $\rmin$ amongst $N$ ABP particles. Normalising by the ideal gas
probability yields $g'(\rmin)=P(\rmin)/P^{\rm ideal}(\rmin)$ whose
asymptotic value at fixed volume
$f_N=\lim_{\rmin\rightarrow\infty}g'(\rmin)$ determines the ratio of
the partition function of $N$ particles to that of an ideal gas. The excess
Helmholtz free energy of the exact effective passive system follows as $-k_BT\ln f_N=F_{ex}$. The method can also be applied to measure $F_{ex}$ for the approximate $2$-body system of $N$ passive particles. The background to our method is described in detail in \cite{ashton2014} and summarised in the SI.}
 
\begin{figure}[t]
\centering
  \includegraphics[width=\columnwidth]{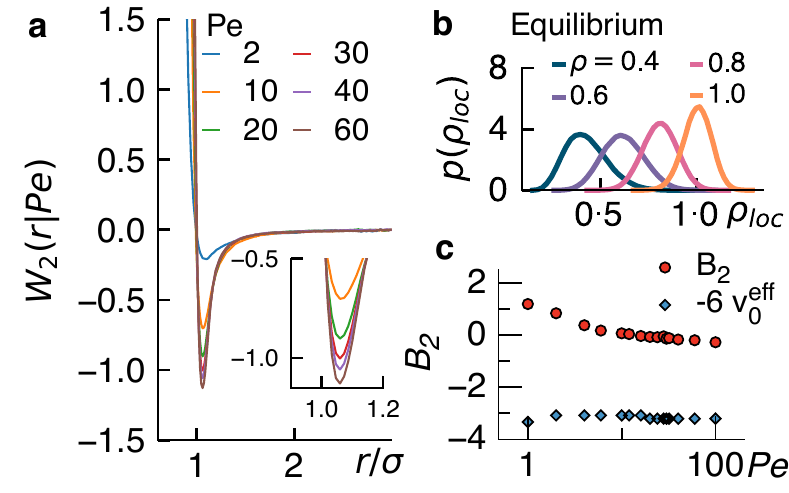}
  \caption{
  (a) The effective pair potential $W_2(r|Pe)$ for several $\Pe$. The inset magnifies the attractive region. (b) Distribution of local density for equilibrium simulations with pair interaction $W_2(r|Pe=60)$ and increasing $\rho$ displaying a single phase. (c) Measured values of $B_2$ together with the criterion for phase separation.}
  \label{fig:twobody}
\end{figure}

The effective pair interaction $W_2(r)$ for $N=2$
ABPs is calculated for a given $Pe$ as $W_2(r)=-\ln \left[g_{2}^{\prime}(r) / f_{2}(V)\right]$,
with $g_{2}^{\prime}(r)= P_{2}\left(r_{\min
}\right)/P_{2}^{\mathrm{ideal}}\left(r_{\min }\right)$ and $f_2 =
\lim_{\rmin\rightarrow\infty}g_2'(\rmin)$ \cite{ashton2014}. The calculation is
repeated for several values of $\Pe$ resulting in the
forms shown in Fig.~\ref{fig:twobody}(a). This
reveals that at low $\Pe$ the interaction is essentially repulsive and
gradually develops an attractive well which becomes deeper as the
$\Pe$ increases. We note that the range of the attraction is short
compared to the size of the repulsive core, in accordance with
equilibrium models having a similar phase diagram topology, and that
the shape of the potential is consistent with analytical
approximations such as the unified coloured noise approximation, known
to reproduce the interactions in the weak activity regime
\cite{wittmann2017a}.

Equipped with the forms of $W(r|Pe)$, we first enquire whether the attraction is sufficient to engender phase separation. Direct simulation with the potential $U(r|\Pe)=k_B T W_2(r|\Pe)$ at $k_B T=1$ demonstrates that this is not the case: for example, in Fig.~\ref{fig:twobody}(b) we show the distribution of local density around the particles for $\Pe=60$ and several total densities well inside the phase separation region, {see Fig.~\ref{fig:phasediagram}}. The distributions are unimodal indicating that no phase separation occurs. 

 The failure of the effective pair potential to yield phase separation can be rationalised by analysing the second virial coefficient \ft{$B_2= -\frac{1}{2}\int(e^{-W_2(r)}-1) d\bm{r}$} and comparing its trends with known criteria for phase separation in equilibrium systems. In simple liquids, {the onset of LV phase separation}  occurs when $B_2 \approx -6 v_0^{\rm eff}$ where $v_0^{\rm eff}$ is the volume of an (effective) hard sphere \cite{vliegenthart2000predicting,noro2000extended}. This {empirical} criterion is particularly accurate in simple liquids with short ranged attractions  \cite{largo2008vanishing,largo2006influence}. However, as shown in Fig.~\ref{fig:twobody}(c), $B_2$ for the effective pair potential for our ABP model never satisfies the criterion even at very large $\Pe$. It follows that the effective, isotropic, two-body interactions obtained via coarse-graining of the orientational degrees of freedom do not engender {sufficiently strong effective} attractions to induce phase separation. \nw{ As shown in the SI \cite{SI}, the same conclusion holds in 2d.}
 
 While one might seek to correct this deficiency by re-introducing the orientational variables and describing the interactions via effective anisotropic short-ranged terms, we find that valuable insight can be gained by refining our calculations to include higher order contributions to the attraction, in the spirit of the multi-body expansions of equilibrium systems \cite{dijkstra1999}. To do so we compare the form of $P_N(\rmin)$ for two systems: the active system at some prescribed $\Pe$ and an approximate passive system with the pair interaction $W_2(r|Pe)$. Specifically, we accumulate the PDFs $P_N^{a}(\rmin)$ of the active and $P_N^p(\rmin)$ of the passive case for $N$ ranging from $N=2$ to $N=24$ in boxes of a fixed volume $V$, such that the number density ranges from $\rho=0.003$ to $0.036\sigma^{-3}$. Comparing active and passive systems of such small $N$ allows us to study systematically the effects of successively higher $N$-body interactions and demonstrate the emergence of collective contributions to the attraction that drives MIPS, similarly to the predictions of effective Fokker-Planck mappings \cite{farage2015}. 

\begin{figure}[t]
\centering
  \includegraphics{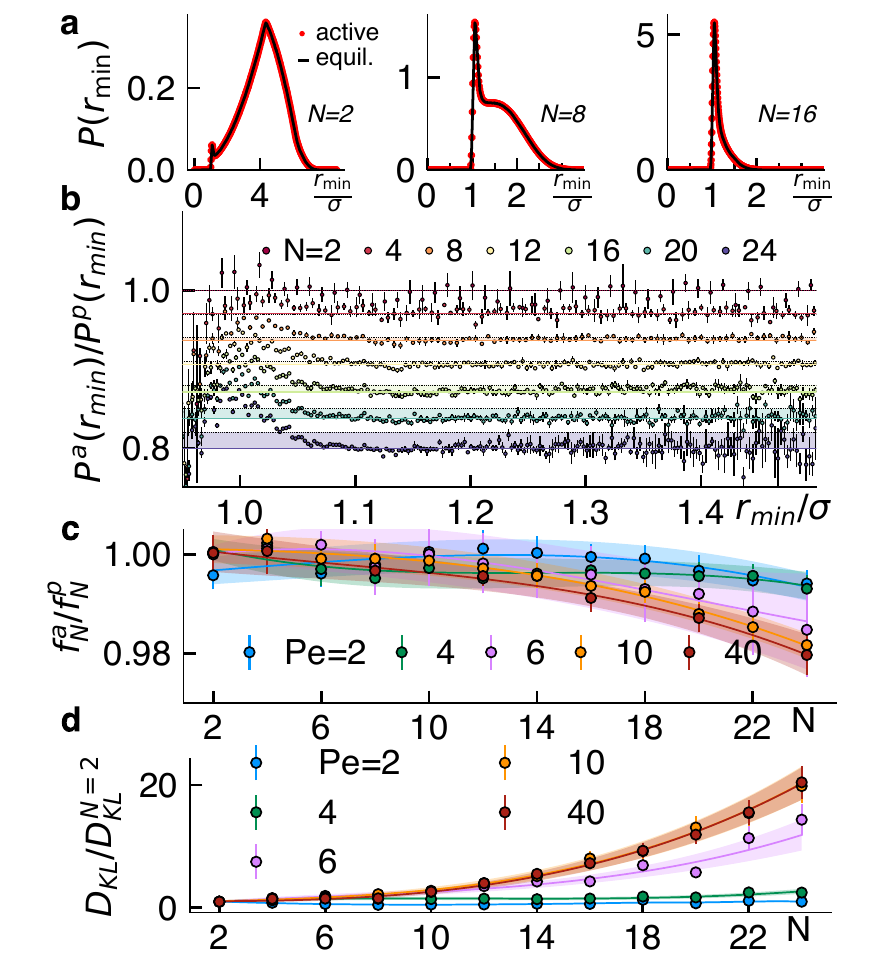}
  \caption{
  (a) $P_N(\rmin)$ for the active system (red circles) and the effective two-body passive system (black line) for three values of $N$ at $\Pe=40$. (b) Ratio $P_N^{a}(\rmin)/P_N^p(\rmin)$ at $\Pe=40$ for increasing $N$. The shaded area indicates the growing gap between $P_N^{a}/P_N^p=1$ and $f_N^{a}/f_N^p$: every curve is shifted vertically for clarity. (c) Ratio between the active and passive asymptotic contributions $f_N$ as estimated from the limit of $P_N^{a}/P_N^p$. (d) {Relative entropy} $D_{KL}$ between active and passive probabilities, scaled by the $N=2$ value. {Lines and shades in (c-d) are polynomial trends and confidence intervals respectively.}}
  \label{fig:multibody}
\end{figure}

Fig.~\ref{fig:multibody}(a) shows that \nw{at first sight} the passive pdf closely reproduces the active ones at $\Pe=40$.  There are nevertheless subtle differences \ft{in the large $\rmin$ regime} that grow with increasing $\Pe$ and which are key to understanding the origin of multi-body effects. To expose these, we consider the ratio $P_N^a(\rmin)/P_N^p(\rmin)$ for various $N$. Results for $\Pe=40$ are shown in Fig.~\ref{fig:multibody}(b) and reveal that on increasing $N$, the active particles are more likely than the passive ones to be found in close contact  -- a fact signalled by a relative depletion of the probability of finding a particle at large distances. The asymptotic value $\lim_{r\rightarrow\infty}P^a(\rmin)/P^p(\rmin)=f^a_N/f^p_N$ is shown in Fig.~\ref{fig:multibody}(c) for a variety of choices of $\Pe$. \nw{Its logarithm measures the free energy difference  between the `exact' passive model of $N$ ABPs and the approximate passive model described by $W(r)$}. While at low $\Pe$, $f^a_N/f^p_N$ is close to unity for all the considered $N$, at sufficiently large $\Pe \gtrsim 6$, the ratio diminishes with increasing $N$. Remarkably, however, the deviation from unity become significant only for $N\gtrsim 12$, i.e. when the number of particles is close to the typical coordination number of a liquid. This suggests that a whole ``cage'' of active particles is needed to engender significant multi-body attractions.

To quantify the difference between the prediction of the passive $2$-body model and the active system, we employ the so-called relative entropy, familiar from information theory, $D_{\mathrm{KL}}(P_N^a \| P_N^p)=\int_{0}^{\infty} P_N^a(\rmin) \log \frac{P_N^a(\rmin)}{P_N^p(\rmin)} d\rmin.$ \cite{kullback1951information}. 
 $D_{\rm KL}$ provides \ft{a scalar measure} of the additional effects (or {\textit{``surprisal''}}) that the model $P^p$ fails to capture \cite{burnham2002practical}. In Fig.~\ref{fig:multibody}(d) we plot $D_{\rm KL}(N)/D_{\rm KL}(N=2)$ for various $\Pe$. At low activity, this ratio is independent of $N$: the two-body passive model provides an accurate representation of the active system. This remains true for $\Pe$ below about $6$, where-after the ratio gradually increases. The increase of $D_{\rm KL}$ is negligible for small $N$: $3$- or $4$-body terms do not contribute significantly to the enhanced attractions. Additionally the behaviour for $\Pe=10$ -- which is just inside the range where enhanced {near}-critical fluctuations are discernible -- is not markedly different from $\Pe=40$. This implies an onset value of P\'eclet number that distinguishes a low activity regime (where multi-body effects are negligible) and a high activity regime \cite{rein2016,fodor2016}.

In conclusion, the phase behaviour of 3d active Brownian particles exhibits striking similarities with that of simple liquids having very short-ranged attractions but it cannot be rationalised \nw{qualitatively} in terms of effective $2$-body interactions. While the pair potentials that we derive in the low density limit are very short-ranged, they fail to yield phase separation. We trace this {fact} to the need to include emergent multi-body terms in the description of the effective model. Whilst in many coarse-grained treatments of equilibrium systems the leading corrections to the pair potential description are given by three and four body interactions\cite{kobayashi2019}, for ABPs the principal multi-body effect that boosts particle attraction and drives MIPS arises when particles become trapped in cages of coordination $\approx 12$ or more.
\footnote{The identified multi-body effects could be incorporated via a density dependent pair potential such as often arise in coarse-grained description of equilibrium systems. However these are known for their inconsistencies \protect\cite{louis2002}}\cite{klongvessa2019}. {Accordingly a mapping of active to equilibrium phase separation can only be achieved at the expense of the simplicity of the equilibrium model.}
\newline

\begin{acknowledgments}
The authors thank R.~Evans, T.~Speck and R.~L.~Jack for insightful conversations and critical reading of the manuscript. This work was carried out using the computational facilities of the Advanced Computing Research Centre, University of Bristol.
\end{acknowledgments}


%

\end{document}